\documentclass[10pt,a4paper]{article}
\usepackage[utf8]{inputenc}
\usepackage{amsmath}
\usepackage{amsfonts}
\usepackage{amssymb}
\usepackage{graphicx}
\usepackage{caption}
\usepackage{subcaption}
\usepackage{epstopdf}
\usepackage{abstract}
\usepackage[toc,page]{appendix}
\usepackage[sort]{cite}
\usepackage{color}
\usepackage{authblk}
\usepackage{placeins}

\newcommand{\half}[0]{\frac{1}{2}}
\newcommand{\bvec}[1]{\mathbf{#1}}

\newcommand{\dd}[0]{\mathrm{d}}

\title{Impact of Interaction Range and Curvature on Crystal Growth of Particles Confined to Spherical Surfaces}
\author[1]{Stefan Paquay*}
\author[1]{Gert-Jan Both}
\author[1,2]{Paul van der Schoot}
\affil[1]{Applied Physics, Technische Universiteit Eindhoven, The Netherlands}
\affil[2]{Instituut voor Theoretische Fysica, Universiteit Utrecht, The Netherlands}
\date{  }

\begin{document}
\maketitle


\begin{abstract}
  When colloidal particles form a crystal phase on a spherical template, their packing is governed by the effective interaction between them and the elastic strain of bending the growing crystal. For example, if growth commences under appropriate conditions, and the circular crystal that forms reaches a critical size, growth continues by incorporation of defects to alleviate elastic strain. Recently it was found experimentally that, if defect formation is somehow not possible, the crystal instead continues growing in ribbons that protrude from the original crystal. Here we report on computer simulations in which we observe both the formation of ribbons at short interaction ranges and packings that incorporate defects if the interaction is longer-ranged. The ribbons only form above some critical crystal size, below which the nucleus is roughly spherically shaped.
  We find that the scaling of the critical crystal size differs slightly from the one proposed by the Manoharan group, and reason this is because the actual process is a two-step heterogeneous nucleation of ribbons on top of roughly circular crystals.

\end{abstract}

\clearpage{}

\section{Introduction}
\label{sec:introduction}


Colloidosomes are droplets whose surfaces are densely packed with colloidal particles. \cite{dinsmore-2002}
Confinement of the colloidal particles to the liquid-liquid interface minimises the contact area between the two liquids.
At high surface coverage, the colloids either pack in a disordered, glassy or in an ordered, crystalline fashion. \cite{dinsmore-2002,sausset-2010,burke-2015,burke-2016}
In the former particles are kinetically trapped.
In the latter the equilibrium packing of the colloids is determined by the interplay between the curvature of the droplet and the exact nature of the interaction between the colloids. This can give rise to various kinds of defect and defect organisations, producing grain boundary scars \cite{bausch-2003,lipowsky-2005,einert-2005}, pleats \cite{irvine-2010} and/or growth in ribbon-like shapes that emanate from circular domains without defects at the boundary between the two. \cite{meng-2014}

Grain boundary scars and pleats involve point defects in specific arrangements, where a point defect is a particle that has fewer or more than six nearest neighbours.
Ribbons form when particles, for whatever reason, refuse to give up local hexagonal order.
One such reason can be elastic stress, resulting from a short interaction range. \cite{meng-2014,koehler-2016}
In macroscopic theory, this stress manifests itself in an additional elastic term in the free energy. \cite{schneider-2005,majidi-2008}
The experimentally observed ribbon-like crystals of Ref. \cite{meng-2014} show an initial, almost isotropic growth that, after a critical size is reached, transitions into the growth of ribbons protruding from the initial crystal. Similar findings are reported in Ref. \cite{koehler-2016}, in which phase field crystal model calculations confirm that growing in ribbons indeed can relieve the elastic stress.
Note that the crystal growth is continuous along the interface between the seed and ribbon, \emph{i.e.}, there is no grain boundary between the two because the ribbon preserves the same lattice vectors.

Theoretical arguments for how the largest circular domain should scale with the Young's modulus and template radius were given by Refs. \cite{meng-2014,koehler-2016} and are based on finding the optimal size of a circular crystal bent onto a spherical surface, which is, in this topology, bounded by an elastic energy penalty.
The theory predicts that ribbons form for a sufficiently short range of interaction, for example of a suitably parametrised Morse potential as employed \emph{e.g.} Ref. \cite{paquay-2016-2}, and only then for sufficiently large domain sizes.

We find by means of computer simulation that the range of the interaction potential between the particles indeed dictates whether or not ribbons form.
For sufficiently large interaction ranges, we find that ribbon formation is suppressed in favour of incorporating defects into the crystal.
For shorter ranges of interaction, the critical size for ribbon formation reduces, which is consistent with Refs \cite{meng-2014,koehler-2016}.
However, the scaling exponent we observe is slightly different, and instead coincides with the scaling of the optimal width of a pure ribbon on a spherical surface proposed in Ref. \cite{grason-2016}.

The remainder of this article is structured as follows.
In Section \ref{sec:methods4} we discuss the simulation and analysis methods.
In Section \ref{sec:results4}, we discuss our computational findings and show that we indeed observe a critical domain size and extract its scaling relations with respect to the potential range and the radius of curvature.
Finally, in Section \ref{sec:conclusion4}, we underline the most important implications of our findings.

\section{Methods}
\label{sec:methods4}
We perform Langevin dynamics (LD) simulations of $N$ particles constrained to a spherical surface with radius $R,$ using the LAMMPS program and a specialised RATTLE algorithm. \cite{plimpton-1995,andersen-1983,paquay-2016}.
The particle number $N$ we couple to the template radius as $Nr_0^2/16 R^2 = 0.4,$ so that if we associate an area $\pi r_0^2/4$ with each particle of diameter $r_0,$ we have the same area coverage $\phi = 0.4$ for all template radii.
Initially the particles are in non-overlapping, random positions on the spherical surface.

We apply a Langevin thermostat with an arbitrary damping time $\tau_L$ to the particles to keep the system at a constant temperature $T$ and to make the particles undergo Brownian motion. $\tau_L$ is the time it takes for the velocity auto-correlation function of a particle to decay to $1/\mathrm{e}$ of its initial value, and is our reference time unit that in effect measures the ratio of the particle's mass and the friction constant. For the interaction potential between particle pairs we use a truncated and shifted Morse potential $U(r)$ 
\begin{equation}
  U(r) = \left[ U_M(r) - U_M(r_c) \right] H(r_c-r),
\end{equation}
where $r$ is the three-dimensional Cartesian inter-particle distance, $r_c$ the cut-off distance, $H(r_c - r)$ the Heaviside step function that is $1$ if $r_c - r > 0$ and $0$ otherwise and $U_M(r)$ is the original Morse potential
\begin{equation}
  U_M(r) = \epsilon \left[ e^{2\alpha(r-r_0)} - 2 e^{\alpha (r-r_0)}\right],
\end{equation}
with a well depth $\epsilon,$ equilibrium spacing $r_0$ and shape parameter $\alpha.$ Throughout the remainder of this article we use $r_0$ as reference length unit and $\epsilon$ as the reference energy unit.

To encourage the growth of a single crystal, we lightly tether the particles to the top of the spherical surface  $(x,y,z) = (0,0,R)$ with a harmonic spring with spring constant $\kappa = 2.5 \epsilon / r_0^2.$ We initialise the structure with this spring in place for $N_s$ time steps at a temperature of $k_B T = 0.3\epsilon,$ after which we remove the springs.
We then equilibrate for another $N_s$ steps while linearly ramping down the temperature from $k_B T = 0.3\epsilon$ to $0.25 \epsilon.$
We verified that a longer ramping time had no influence on the formed structures.
After this annealing phase we sample for $N_s$ time steps at $k_B T = 0.25\epsilon.$
We analyse snapshots that are $5~\tau_l$ apart and average over them to get statistics.
For $\alpha = 40/r_0$ a smaller time step of $0.001 \tau_L$ was required and we set $N_s =  25\times10^6$ to guarantee proper equilibration and stable dynamics, while for the other values of $\alpha$ we used $N_s = 250\times10^3$ and a time step size $0.005\tau_L.$

\begin{figure}
  \centering
  \includegraphics[width=0.8\linewidth]{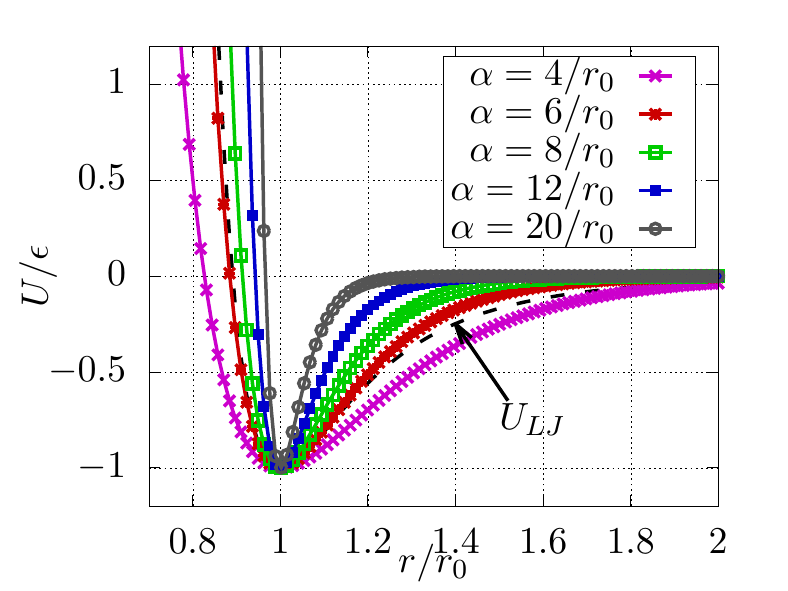}
  \caption{Comparison of Morse (coloured lines/symbols) and Lennard-Jones (dashed black) potentials. Increasing $\alpha$ leads to a sharper well for the Morse potential. For $\alpha = 6/r_0$ the harmonic approximation for the Morse potential equals that of the Lennard-Jones potential.
  \label{fig:LJ-and-morse2}}
\end{figure}

The parameter $\alpha$ sets how sharp the potential is peaked around its minimum at $r_0$, as the effective spring constant $k = \partial^2U(r)/\partial r^2|_{r_0} = 2 \alpha^2 \epsilon.$
See Fig. \ref{fig:LJ-and-morse2} for a graph comparing the Lennard-Jones and Morse potentials for various values of $\alpha.$
If we ignore the impact of multi-body interactions, the Young's modulus scales linearly with $k.$ This implies that a larger $\alpha$ leads to a larger Young's modulus \cite{vitelli-2006}.
Therefore, in our computer simulations we can use the parameter $\alpha$ to influence the Young's modulus of our crystal. By studying the particle vibrations about their lattice site, however, we see that the pair approximation is not very accurate.
Hence, we rely on the actual vibrations to estimate the effective spring constant.
See Section S1 of the supporting material for a detailed description of the analysis.

For our crystals it turns out that effective spring constants $\kappa$ scale significantly less strongly with $\alpha$ than predicted by the harmonic approximation $\kappa \sim \alpha^2.$
We obtain scalings between $\kappa \sim \alpha^{1.3}$ to $\kappa \sim \alpha^{1.6},$ with an average exponent of $(1.52\pm 0.08).$
The deviation from the pair potential approximation could be either due to collective effects that are obviously not included in any pair-wise approximation, or due to the non-zero temperature in our simulations, which does influence elastic properties. \cite{squire-1969}
This is not a complete surprise as hard sphere crystals have an effective spring constant that follows entirely from many-body effects. \cite{lowen-1990}
This shows that it is critical to measure Young's modulus or at least the effective spring constant of the material at hand rather than relying on the harmonic approximation  \cite{vitelli-2006,koehler-2016} when performing Langevin dynamics simulations.
With the effective spring constant determined, we still need methods to determine the scaling of the chemical potential and the largest circular domain size with $\alpha.$


For larger values of $\alpha$ at which the crystal forms ribbons, most of the direct interaction is governed by nearest-neighbour interactions, as can be inferred from Fig. \ref{fig:LJ-and-morse2}.
Because of this, we assume the line tension arises predominantly from the edge particles having fewer than six bonds.
In this case, we expect that the line tension and the chemical potential scale similarly with $\alpha,$ and we only have to determine $\Delta \mu.$

To determine the chemical potential, we have to determine the free energy difference between a crystal consisting of $N$ Morse particles and that same crystal in which one of the bulk particles is changed to an ideal gas particle.
This is problematic because an ideal gas particle tends to explore all of the spherical surface, while the Morse particle tends to stay near its lattice site.
Therefore, we instead determine the free energy difference associated with transforming one Morse particle into an ``Einstein particle'', that is, a particle that is tethered to its lattice site by a harmonic spring.

As reference Morse particle we take the particle that has the largest minimum distance to the edge particles.
By applying Bennett's acceptance ratio \cite{bennett-1976} to the case of $N$ Morse particles and the case of where we transform this reference particle into an Einstein particle, we obtain free energy differences that we can directly relate to the excess chemical potential.
In Supporting Material S2 a more detailed description of the procedure is presented.
The found scaling is $\Delta \mu \sim \alpha^{(-0.53 \pm 0.01)}.$
Note that the free energy difference goes through an appreciable range as a function of $\alpha,$ as we find $\Delta \mu \approx -0.7/\epsilon$ for the largest $\alpha r_0 = 40,$ which is a third of that of $\Delta \mu \approx -2.2/\epsilon$ for the smallest $\alpha r_0 = 4.$
It is thus important to take this scaling into account when assessing the scaling for the largest circular domain size at which the growth transitions to ribbons.

With the scalings for the effective spring constant $\kappa$ and hence Young's modulus $Y$ and that of the chemical potential $\Delta\mu$ and line tension $\gamma$ determined, we can now determine how the largest circular domain sizes scale as a function of the aforementioned parameters, provided that we can extract what the largest circular domain size is.
We describe our approach now in brief.

Finding the largest circular domain size involves identifying the edge of what constitutes a circular domain and extracting the distance from that edge to the centre of the domain, which can be reasonably approximated by finding the particle that is the furthest away from all edge particles.
We calculate for all particles the shortest distance to the edge.
The particle that is the farthest away from the edge we then consider to be the centre of a circular domain, and its distance to the edge, divided by the template radius $R,$ we take as value for the diameter of the largest circular domain.
Note that this particle coincides to that we determine the chemical potential for.

Incidentally, this value also gives a good indication of the transition from incorporated point defects to ribbons, as this number scales differently with the range parameter $\alpha$ when the crystal incorporates point defects, as we shall see in Section \ref{sec:results4}.
This analysis we perform with post-processing scripts that are described in more detail in Supporting Material S3.
With all the methods discussed, we now present our results.

\section{Results}
\label{sec:results4}
\begin{figure}[tb!]
  \begin{center}
        \begin{subfigure}{0.38\textwidth}
      \includegraphics[width=\textwidth]{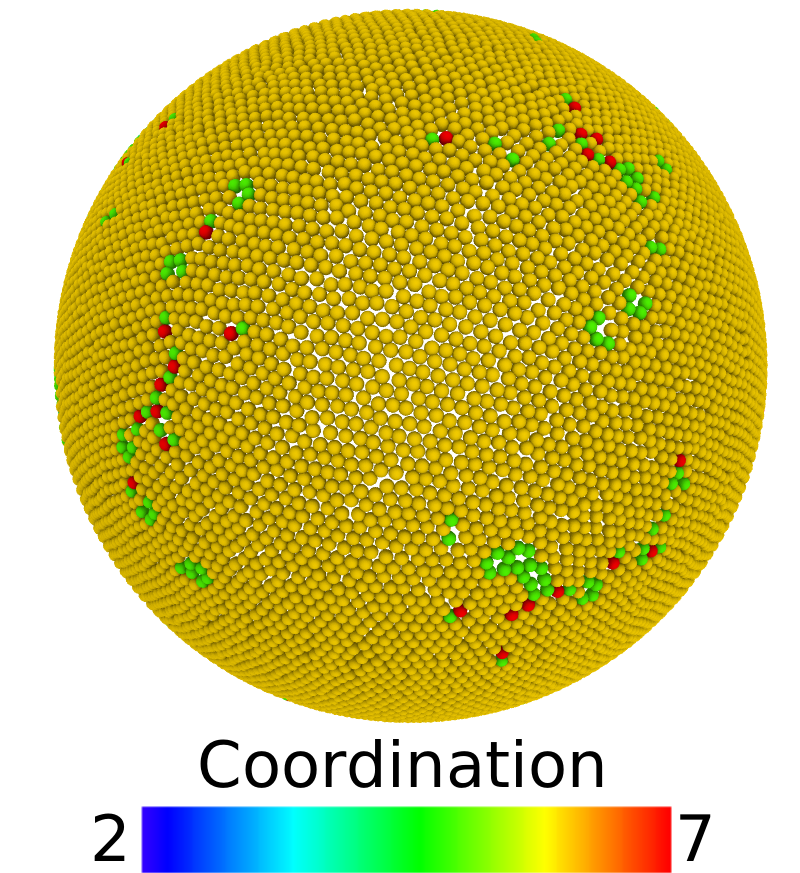}
      \caption{$\alpha r_0 = 4,~R/r_0 = 30.$ \label{}}
    \end{subfigure}
    \begin{subfigure}{0.38\textwidth}
      \includegraphics[width=\textwidth]{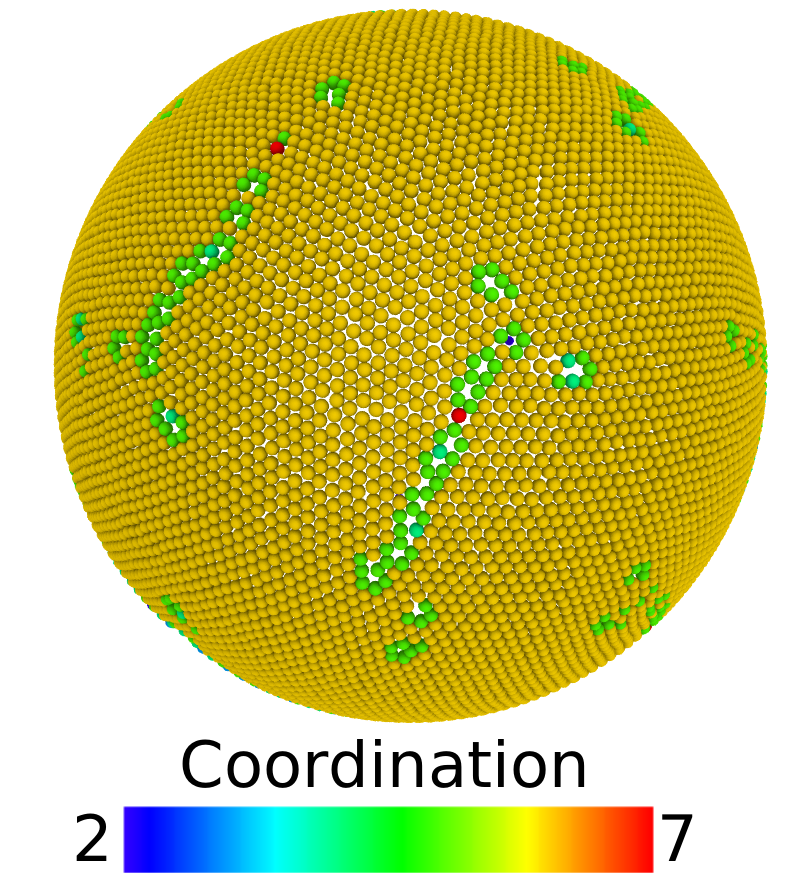}
      \caption{$\alpha r_0 = 8,~R/r_0 = 30.$ \label{}}
    \end{subfigure} \\
    \begin{subfigure}{0.38\textwidth}
      \includegraphics[width=\textwidth]{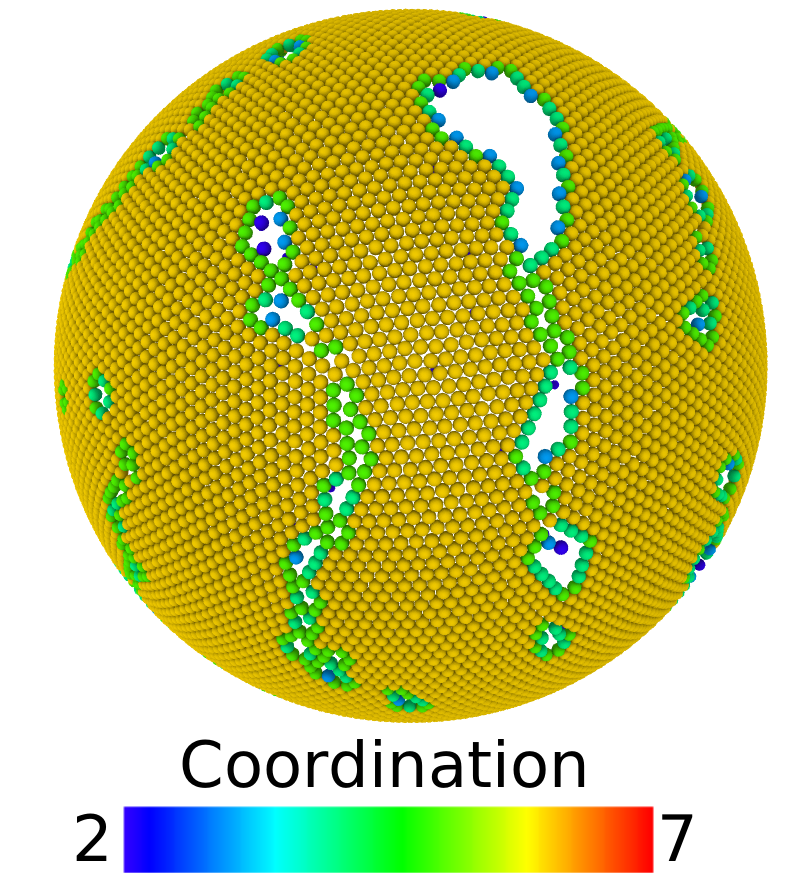}
      \caption{$\alpha r_0 = 16,~R/r_0 = 30.$ \label{}}
    \end{subfigure}
    \begin{subfigure}{0.38\textwidth}
      \includegraphics[width=\textwidth]{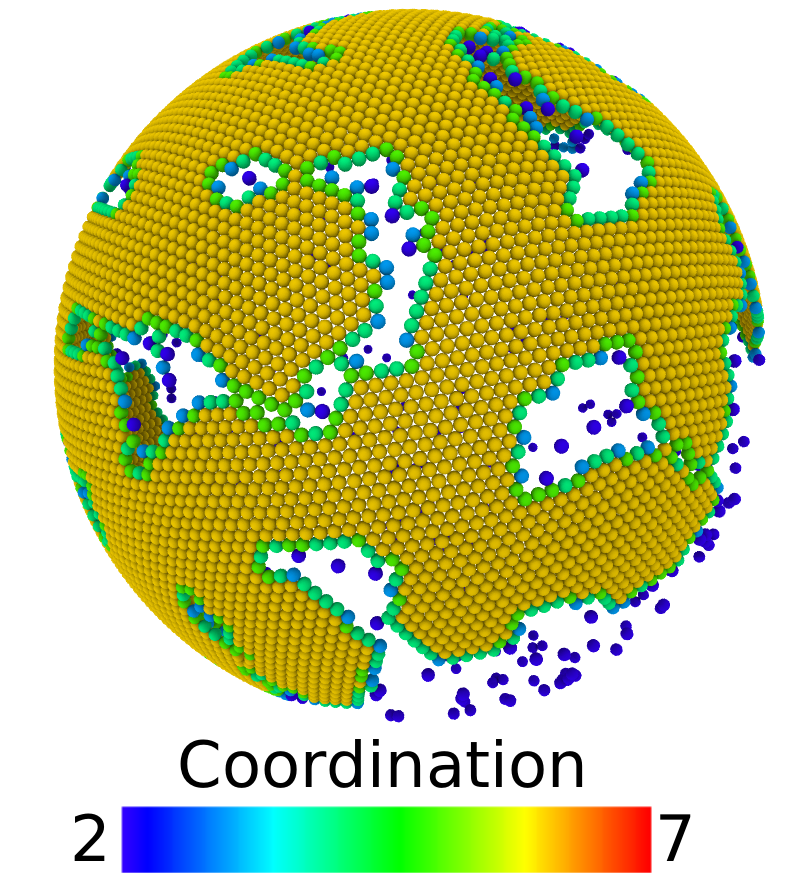}
      \caption{$\alpha r_0 = 32,~R/r_0 = 30.$ \label{}}
    \end{subfigure}
    \caption{Top view of typical crystals obtained from the simulation protocol at an area coverage of $40\%.$
      Colour codes for the number of nearest neighbours (coordination). Images rendered by Ovito \cite{ovito}.
      The back of the spherical template is not covered except by particles in a gaseous phase that forms for $\alpha r_0 \geq 16.$
      Note that for $\alpha r_0 = 4$ the crystal incorporates dislocations and point defects in the structure, while for $\alpha r_0 \geq 8$ tears and holes form between patches of hexagonal lattice.
    \label{fig:typical-crystals}}
  \end{center}
\end{figure}

In Fig. \ref{fig:typical-crystals} we present typical crystal structures we observe as a function of the range parameter $\alpha.$
The figures show the top half of the spherical template that is covered with particles.
For  $\alpha r_0 = 4$ the crystal incorporates various defects, predominantly dislocations (Fig. \ref{fig:typical-crystals}(a)).
At around $\alpha r_0 = 8$ a transition from incorporated defects to hexagonal packings occurs.
From this point on the defects are only located at the edge or around holes (Fig. \ref{fig:typical-crystals}(b)).
Increasing $\alpha$ further to $\alpha r_0 = 16$ (Fig. \ref{fig:typical-crystals}(c)) and $\alpha r_0 = 32$ (Fig. \ref{fig:typical-crystals}(d)) leads to a clear formation of ribbon-like structures separated by larger tears.
Furthermore the width of the ribbons reduces, which is qualitatively consistent with both the scalings presented in Refs. \cite{meng-2014,koehler-2016} and Ref. \cite{grason-2016}.

Note finally that at larger $\alpha$ the effective temperature of the particles appears to have increased slightly, in the sense that more particles appear in a gas-like phase on the side of the template that is not covered by the crystal.
The motion of the particles inside the crystal, however, becomes smaller due to the increased sharpness of the potential well depth.

Our crystals are reminiscent of the ones observed in simulations by Cong \emph{et al.} in Ref. \cite{cong-thesis}.
For $\alpha r_0 \geq 8$ we observe hexagonal patches separated by tears and holes, \emph{i.e.}, ribbons.
Below that $\alpha$ the crystal incorporates pleats and point defects, which are reminiscent of those observed experimentally in Refs. \cite{bausch-2003,lipowsky-2005,einert-2005,irvine-2010}.
Hence, as shown experimentally in Ref. \cite{meng-2014} and consistent with the calculations of Ref. \cite{koehler-2016}, ribbons only form when the elastic strain, regulated by the range of the potential, prevents the formation of point defects.
Interestingly, we do not seem to observe the four-fold branching described in Ref. \cite{koehler-2016}, possibly due to the thermal fluctuations in our simulations, which are absent in a phase-field crystal model.

\begin{figure}[tb!]
  \begin{center}
    \begin{subfigure}{0.324\textwidth}
      \includegraphics[width=\textwidth]{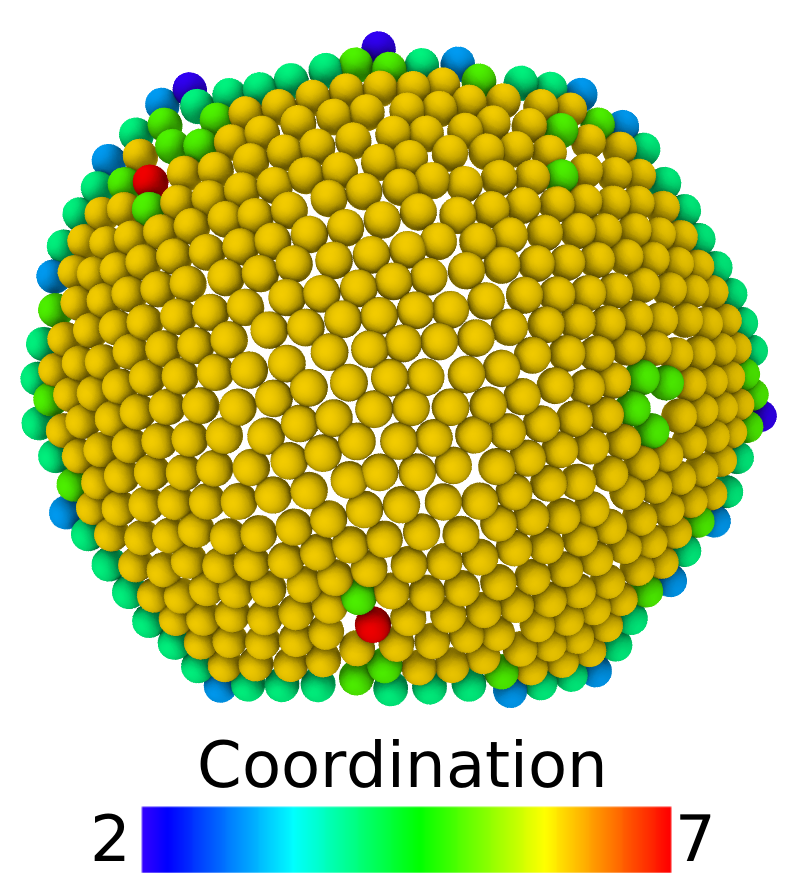}
      \caption{$\alpha r_0 = 4$}
    \end{subfigure}
    \begin{subfigure}{0.324\textwidth}
      \includegraphics[width=\textwidth]{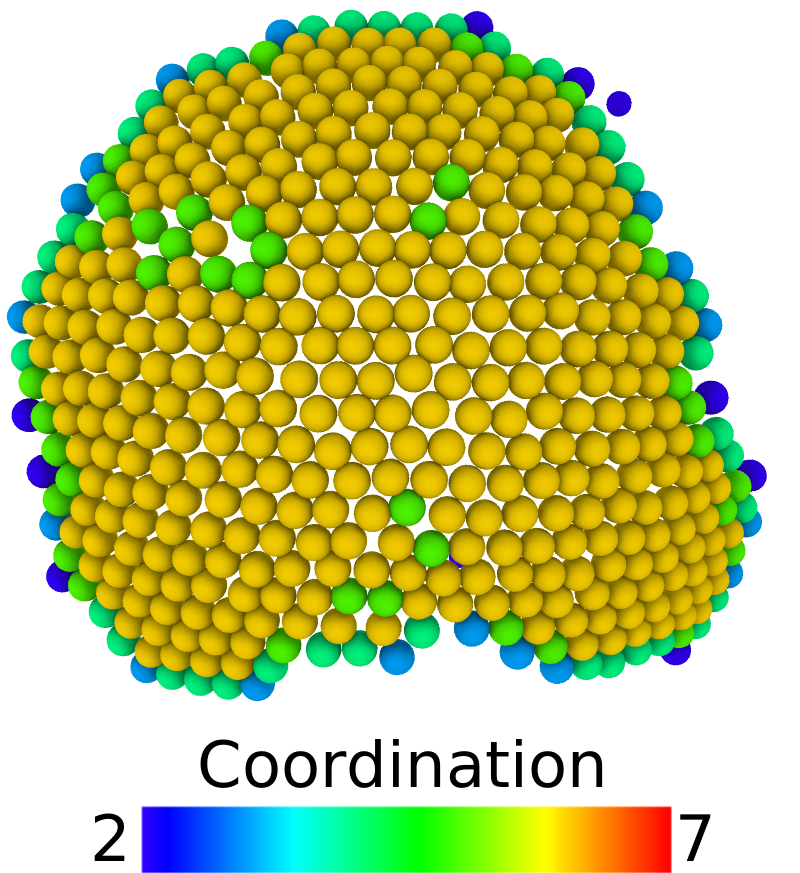}
      \caption{$\alpha r_0 = 6$}
    \end{subfigure}
    \begin{subfigure}{0.324\textwidth}
      \includegraphics[width=\textwidth]{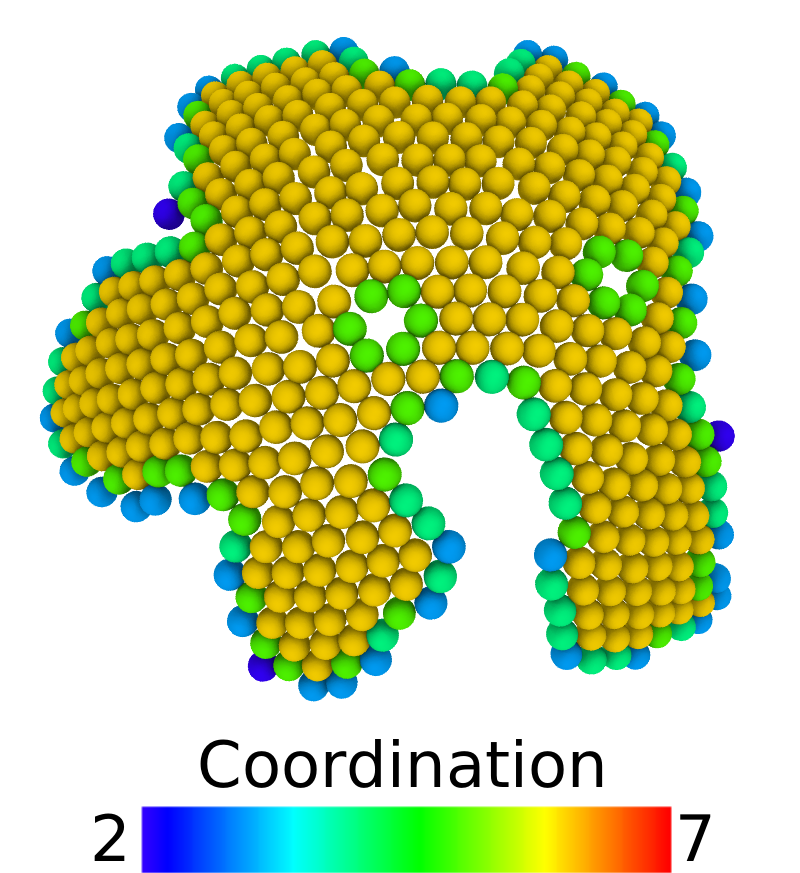}
      \caption{$\alpha r_0 = 8$}
    \end{subfigure}
    \caption{Top view of crystals illustrating the transition from incorporated defects (a) through an intermediate (b) to a predominantly branched structure (c) for 450 Morse particles interacting with varying $\alpha$ at an area coverage of $12.5\%.$
      \label{fig:ribbon-vs-defects}}
  \end{center}
\end{figure}

In Fig. \ref{fig:ribbon-vs-defects} we show an illustrative example of the difference between ribbons and incorporated point defects, obtained by applying the protocol of Section \ref{sec:methods4} to $N=450$ particles on a template of radius $R = 15r_0$ for $\alpha r_0 = 4,~6$ and $12.$
This example is obtained at an area coverage to $12.5\%$ to illustrate more clearly the different structures.
For $\alpha r_0 = 4$ the crystal is roughly spherical and incorporates some defects.
For $\alpha r_0 = 6$ we see the onset of a ribbon that is branching out of the initial bulk while there are still defects in the original nucleus as well.
Hence, there appears to be a sort of ``coexistence'' between incorporated defects and ribbon formation.
For larger $\alpha r_0 = 8$ we see clear branching, and the defects in the bulk are significantly reduced.

\begin{figure}[tb]
  \begin{center}
    \begin{subfigure}{0.48\textwidth}
      \includegraphics[width=\textwidth]{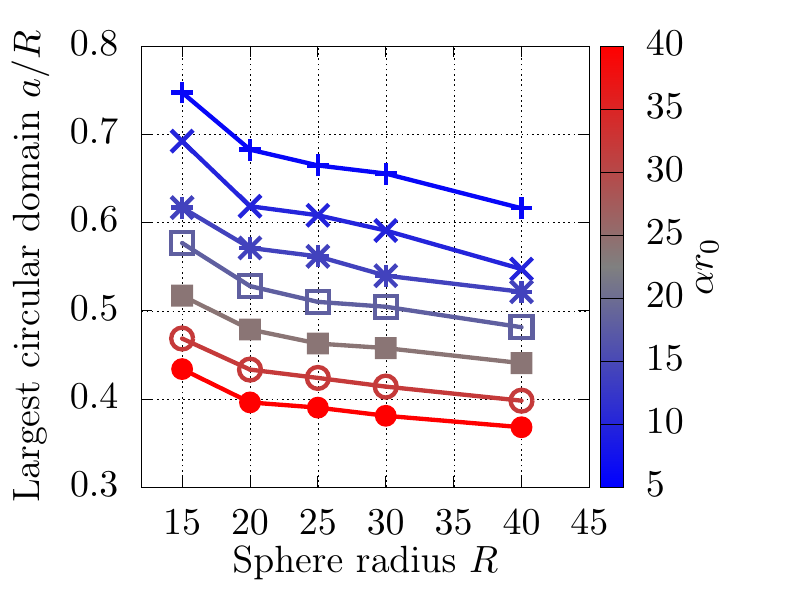}
      \caption{\label{}}
    \end{subfigure}
    \begin{subfigure}{0.48\textwidth}
      \includegraphics[width=\textwidth]{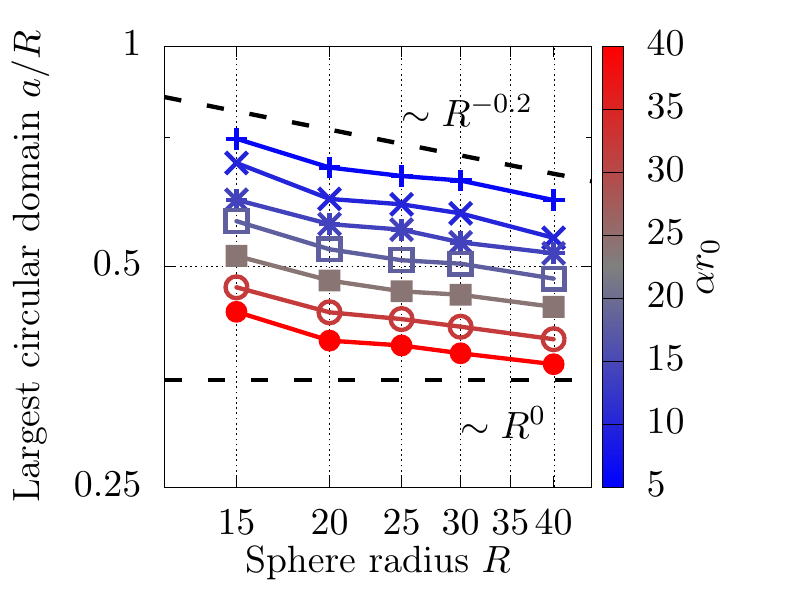}
      \caption{\label{}}
    \end{subfigure}
    \caption{Scaling of the critical circular domain diameter $a$ with the template radius $R.$
      Both the linear (a) and log-log scale (b) reveal that $a$ decreases with increasing $R.$
      \label{fig:scale-R}}
  \end{center}
 \end{figure}

We now analyse the size of the largest circular domain we observe.
All simulation results from which the largest circular domain size is extracted below correspond to a higher area coverage of $40\%$ rather than the $12.5\%$ that served as an illustration of the morphology in Fig. \ref{fig:ribbon-vs-defects}.
First we consider its scaling with the spherical template radius.
Figures \ref{fig:scale-R}(a) and \ref{fig:scale-R}(b) reveal that the critical domain size diameter $a$ scales sublinearly with $R.$
Fitting a power law reveals the scaling exponent to be $(0.83 \pm 0.06),$ which is inconsistent with the scaling proposed in \cite{meng-2014,koehler-2016}.

It does match the scaling of the optimal ribbon width proposed in Ref. \cite{grason-2016}, as well as our own theory based on heterogeneous nucleation of a ribbon-like structure on a circular nucleus, presented in the Supporting Material S4.
Note however that the two theories are very similar, predicting either $a \sim R (\rho \Delta \mu/Y)^{1/4}$ or $a \sim R (\gamma/RY)^{1/5}.$
Hence, while our result quantitatively agrees more closely with the two-stage nucleation model, simulations on larger templates are required to give a definitive answer.

\begin{figure}[tb]
  \begin{center}
    \begin{subfigure}[t]{0.48\textwidth}
      \includegraphics[width=\textwidth]{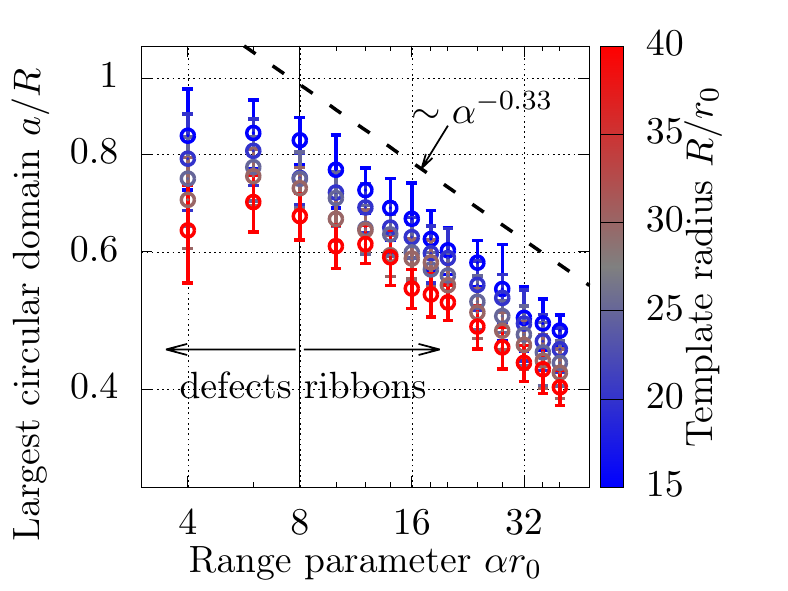}
      \caption{\label{fig:scaling-alpha-linear}}
    \end{subfigure}
    \begin{subfigure}[t]{0.48\textwidth}
      \includegraphics[width=\textwidth]{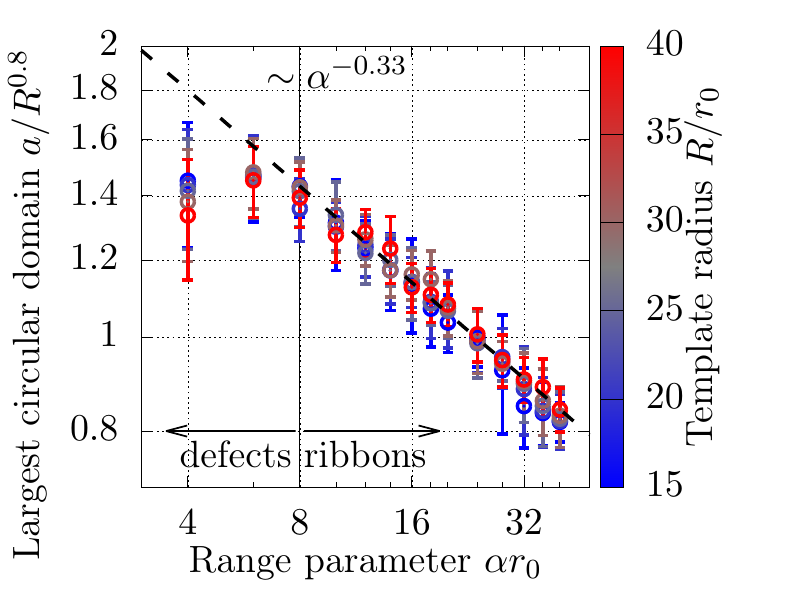}
      \caption{\label{fig:scaling-alpha-loglog}}
    \end{subfigure}
    \caption{Scaling of the critical domain size $a$ with the potential shape parameter $\alpha$ (a)
      at a constant area coverage of $40\%.$
      Dividing out the scaling with $R^{-0.2}$ collapses all data onto one line (b).
      The scaling converges to $a \sim \alpha^{-0.33}$ for large $\alpha.$
      The transition from incorporated defects to ribbons at around $\alpha r_0 = 8$ is apparent from the different scaling with $\alpha.$
      \label{fig:scaling-alpha}}
  \end{center}
\end{figure}

The figures also show that the largest circular domain size decreases with increasing $\alpha.$
This is qualitatively consistent with either theory.
In Figs. \ref{fig:scaling-alpha} we plot the observed largest domain diameter divided by the scaling argument presented in Refs. \cite{meng-2014,koehler-2016} (Fig. \ref{fig:scaling-alpha-linear}) and the heterogeneous nucleation model of Supporting Material S4 (Fig. \ref{fig:scaling-alpha-loglog}).

For $\alpha r_0 \leq 6,$ where the crystal incorporates point defects, the domain size appears to be independent of $\alpha r_0.$
This makes sense as the largest circular domain will only depend on the number of particles in the crystal, which, for our simulation setup, scales linearly with $R.$
Since the scaling of the largest domain size with $\alpha$ is clearly not relevant in this regime, we omit the data for $\alpha r_0 < 8$ in the following analysis of the largest circular domain size.

Note that the data presented in Fig. \ref{fig:scaling-alpha}(b) collapses onto a single curve, giving a strong indication that in the parameter range we have access to, two-stage nucleation is the main pathway for ribbon formation.
For $\alpha r_0 \geq 8,$ fitting a power law gives us access to the second scaling exponent of $(-0.332\pm 0.005),$ which, in combination with the scaling exponent with $R$ of $(0.83 \pm 0.06),$ quantifies our observed scaling of $a \sim R^{(0.83)} \alpha^{-0.33}.$
The heterogeneous nucleation model predicts, with our empirical scaling of $Y$ and $\gamma$ with $\alpha,$ that the second scaling exponent should be $(-0.41\pm0.02),$ whereas the original scaling proposed in Refs. \cite{meng-2014,koehler-2016} leads to an exponent of $(-0.51 \pm 0.02).$
Again, the heterogeneous nucleation model seems to better fit our observations, although the scaling with the interaction range is certainly less convincing than the scaling with the spherical template radius $R.$

\section{Discussion and conclusion}
\label{sec:conclusion4}
We performed Langevin dynamics simulations of Morse particles of diameter $r_0$ on spherical surfaces of varying radius.
We can tune the range of interaction between the particles with a parameter $\alpha,$ where larger values of $\alpha$ represent shorter ranges of attraction.
We only consider a single interaction strength $\epsilon = 4k_BT,$ at which we obtained well-equilibrated crystals.
Below a critical $\alpha r_0 \leq 6$ the crystals that form are roughly spherical and incorporate dislocations and point defects.
Above this critical value the crystals instead exhibit a smaller, defect-free, circle-like nucleus with protruding, ribbon-like structures.
This provides a confirmation of the suggestion that the formation of the ribbons is indeed driven by the elastic instability put forward in Ref. \cite{meng-2014}.

We quantified the scaling of the size of these circular domains as a function of both the template radius $R$ and the range parameter $\alpha$ and find that the largest circular domain diameter scales as $a\sim R^{(0.83\pm0.06)}\alpha^{(-0.33\pm0.01)}.$
This is close to but not consistent with the scaling predicted in Refs. \cite{meng-2014,koehler-2016}, which assume the critical circular domain size follows from the global minimum in the free energy due to the elastic penalty.
Rather, we find that the scaling is more consistent with a heterogeneous nucleation model, where the transition is not dictated by the global minimum in the free energy but rather by the fact that for a sufficiently large circular domain, continued growth as a ribbon is energetically more favourable than as a circle.
However, we varied both parameters through at most one decade, so further studies on larger crystals are required to determine the full parameter range over which the model is valid.

Finally, more generally, our findings indicate that, in addition to its well-known influence on the number of local minima and liquid phase stability in free space, \cite{doye-1995,doye-1996,doye-1996-2,hagen-1993} the range of attraction also influences the morphology of the formed crystal on a curved surface.
This has important consequences for, \emph{e.g.}, the formation of virus capsids, whose building blocks interact through short-ranged interactions \cite{parsegian-boek} but are typically modelled by a long-ranged Lennard-Jones potential \cite{zandi-2004}.

\section{Acknowledgements}
We are grateful to Mike Hagan, Greg Grason, Jayson Paulose, Vinothan Manoharan and Wouter Ellenbroek for valuable discussions, and Thijs van der Heijden for critical proof-reading.
We acknowledge the HFSP for funding under grant RGP0017/2012.


{

\renewcommand{\thetable}{S\arabic{table}}
\renewcommand{\thesection}{S\arabic{section}}
\renewcommand{\thefigure}{S\arabic{figure}}
\setcounter{figure}{0}
\setcounter{table}{0}
\setcounter{section}{0}

\clearpage{}

\begin{center}
  {\LARGE{Supporting Material for Impact of Interaction Range and Curvature on Crystal Growth of Particles Confined to Spherical Surfaces}}
\end{center}

\section{Extracting effective spring constants}
\label{sec:si1}
In order to estimate the scaling of Young's modulus, we require the the effective spring constant $\kappa$ that keeps the particles in place in the crystal.
For a single particle pair this can be found analytically by harmonically approximating the pair potential around the minimum.
For the case of the Morse potential, this leads to $\kappa = 2\alpha^2\epsilon.$
However, this simple approximation does not take into account  collective effects.
Therefore, we extract effective spring constants from our simulation data.

From our simulation trajectories, we determine a running average of the particle positions $\bvec{x}$ by applying an exponentially weighted average to obtain for a time step $t_n$ the average $\bvec{x}_a(t_n) = (1 - \xi)\bvec{x}(t_n) + \xi\bvec{x}_a(t_{n-1}).$
The parameter $\xi$ controls how fast the past positions are ``forgotten''.
At each time step, we determine the squared deviation of the particles from the average from the previous frame, $\delta \bvec{x}^2(t_n) := (\bvec{x}(t_n) - \bvec{x}_a(t_{n-1}))^2.$
This serves as an approximation of the squared deviation of the particle from its average lattice site while taking into account that the lattice site might drift in time.
Typical squared displacement values we find are of the order of $0.05 r_0^2.$
We empirically tuned $\xi$ to a value of 0.9 at which the fast motions about the average lattice site are suppressed.

This produces for each particle a time trace of the quantity $\delta \bvec{x}^2,$ which can be averaged in time to obtain $\left< \delta \bvec{x}^2\right>.$
If we assume particles are bound to their lattice site by a harmonic spring, the effective spring constant then follows from equipartition as $\half \kappa \left< \delta \bvec{x}^2\right> = k_B T.$
In Fig. \ref{fig:histograms-effective-k} we present distributions for the squared deviation and effective spring constant for four values of $\alpha.$

In Fig. \ref{fig:effective-ks} we present the effective spring constants for all parameter values. Note that for very large $\alpha$ the scaling breaks down.
This is because the script used to determine the peak in the histogram simply determines the numerical maximum as an approximation for the mean.
This breaks down for noisy and particularly wide distributions, which we obtain for large $\alpha,$ as can be seen from the grey and red distributions in Fig. \ref{fig:histograms-effective-k}(b).

\begin{figure}[tb]
  \begin{center}
    \begin{subfigure}{0.48\textwidth}
      \includegraphics[width=\textwidth]{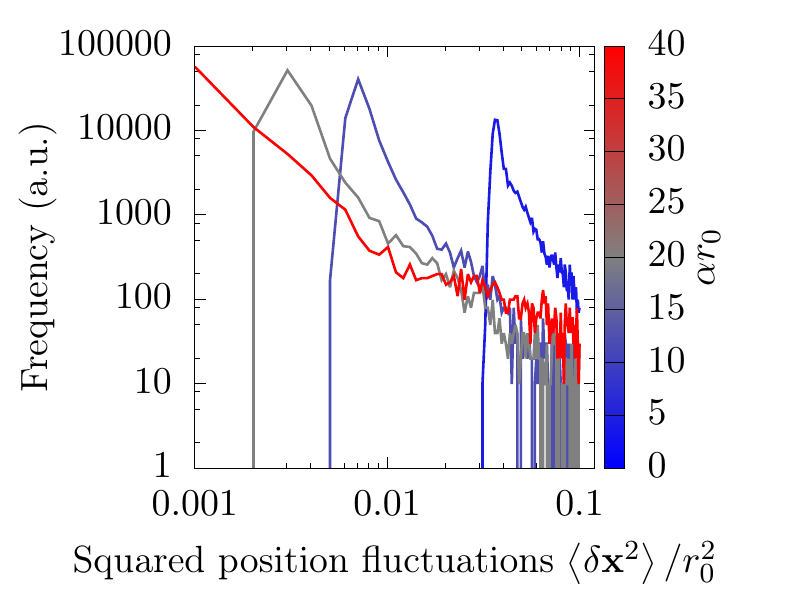}
      \caption{\label{}}
    \end{subfigure}
    \begin{subfigure}{0.48\textwidth}
      \includegraphics[width=\textwidth]{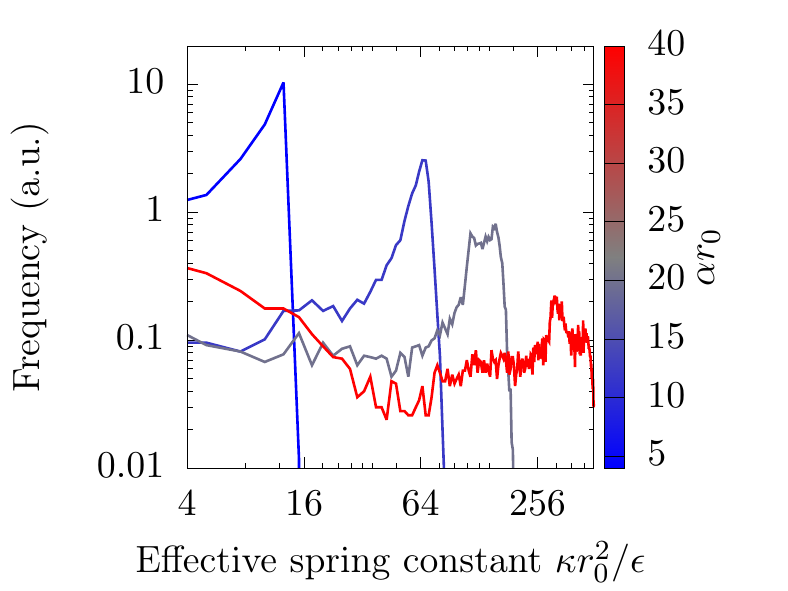}
      \caption{\label{}}
    \end{subfigure}
    \caption{Distributions of the squared fluctuations in the particle positions about their lattice site (a) and of the effective spring constant $\kappa = 2k_BT/\left< \delta \bvec{x}^2 \right>$ (b), both for $R = 40~r_0$ and an area coverage of $40\%.$
      For increasing $\alpha$ the peak in $\left< \delta \bvec{x}\right>^2$ shifts to the left and, consistently, the peak in $\kappa$ shift to the right.
      \label{fig:histograms-effective-k}}
  \end{center}
\end{figure}

\begin{figure}[tb]
  \begin{center}
    \includegraphics[width=0.6\textwidth]{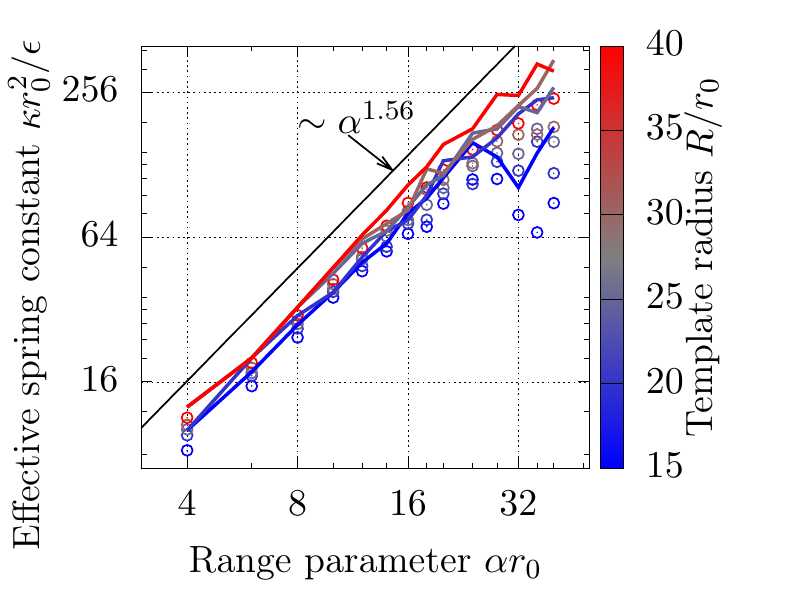}
    \caption{Extracted modal (lines) and average (symbols) of the effective spring constant distributions from Fig. \ref{fig:histograms-effective-k}.
      The data for $R=40~ r_0$ best follows the power law $\kappa \sim \alpha^{1.56}$ with the asymptotic standard error of the fit being $0.05.$ \label{fig:effective-ks}}
  \end{center}
\end{figure}

\section{Determining the chemical potential}
\label{sec:si2}
To determine the chemical potential of a particle in the crystal, we determine the free energy difference between the original crystal of $N$ Morse particles, and a hybrid Morse-Einstein crystal in which the particle which is most representative of the bulk of the crystal is replaced by an Einstein particle tethered to its initial lattice site with a harmonic spring.
To determine the most representative particle, we apply the analysis to extract the largest circular domain size, which we shall discuss later in Appendix  \ref{sec:si3}.
This analysis finds the particle that is the furthest away from all edge particles, which is the best available representative of the bulk.

We put the spring constant to $k=50 \epsilon/r_0^2,$ at which the Einstein particle and the original Morse particle have comparable root mean square deviations from the average lattice site, calculated following the procedure detailed in Appendix \ref{sec:si1}.
We apply Bennett's acceptance criterion \cite{bennett-1976} to the two different cases, whose total potential energies are
\begin{align}
  U_0 =& \sum_{i=0}^{N-1}\sum_{j>i}^{N-1} U(r_{ij}) + \sum_{i=0}^N U(\bvec{r}_{iN}),\qquad U_1 = \sum_{i=0}^{N-1}\sum_{j>i}^{N-1} U(r_{ij}) + U_E(\bvec{x}_N),\\
  r_{ij} :=& \left\| \bvec{x}_i - \bvec{x}_j\right\|, \qquad U(r) = \left[U_M(r) - U_M(r_c)\right]H(r_c-r), \\
  U_M(r) :=& \epsilon\left[e^{-2\alpha(r-r_0)} - 2e^{-\alpha(r-r_0)}\right], \qquad U_E(\bvec{x}_N) = \frac{1}{2} \kappa ( \bvec{x}_N - \bvec{x}_N(0))^2.
\end{align}
Here, $U_0$ is the original truncated, shifted Morse potential $U$ with cut-off distance $r_c,$ well depth $\epsilon$ and range parameter $\alpha$ acting on all $N$ particles with diameter $r_0,$ while $U_1$ is the original Morse potential acting on $N-1$ particles combined with an ``Einstein potential'' $U_E$ applied to the $N$th particle.
$H(x)$ is the Heaviside function with $H(x) = 1$ for $x > 0$ and $0$ otherwise.
The Einstein potential is, of course, a harmonic spring with spring constant $\kappa$ that attaches the $N$th particle to its initial position $\bvec{x}_N(0).$
Finally, $r_{ij}$ is the scalar distance between particles $i$ and $j,$ located at positions $\bvec{x}_i$ and $\bvec{x}_j.$

\begin{figure}[tb]
  \begin{center}
    \begin{subfigure}[t]{0.48\textwidth}
      \includegraphics[width=\textwidth]{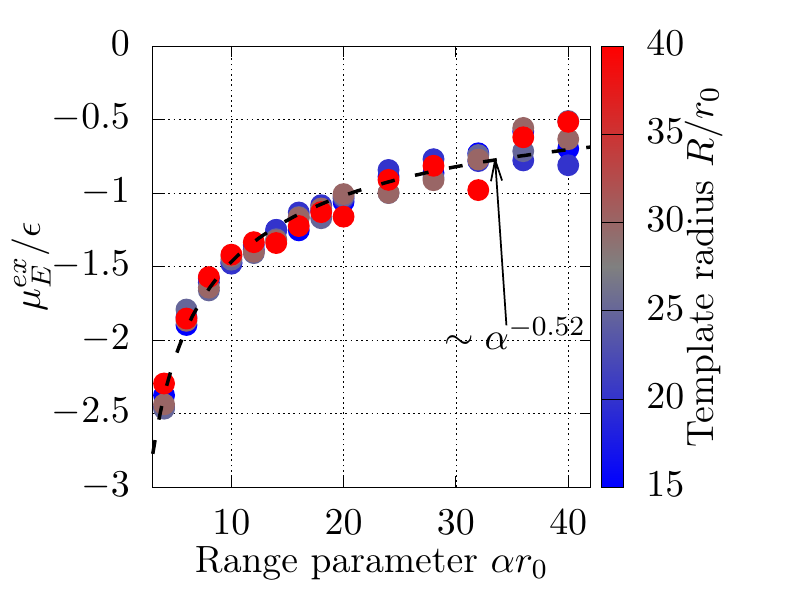}
      \caption{\label{fig:scaling-alpha-linear}}
    \end{subfigure}
    \begin{subfigure}[t]{0.48\textwidth}
      \includegraphics[width=\textwidth]{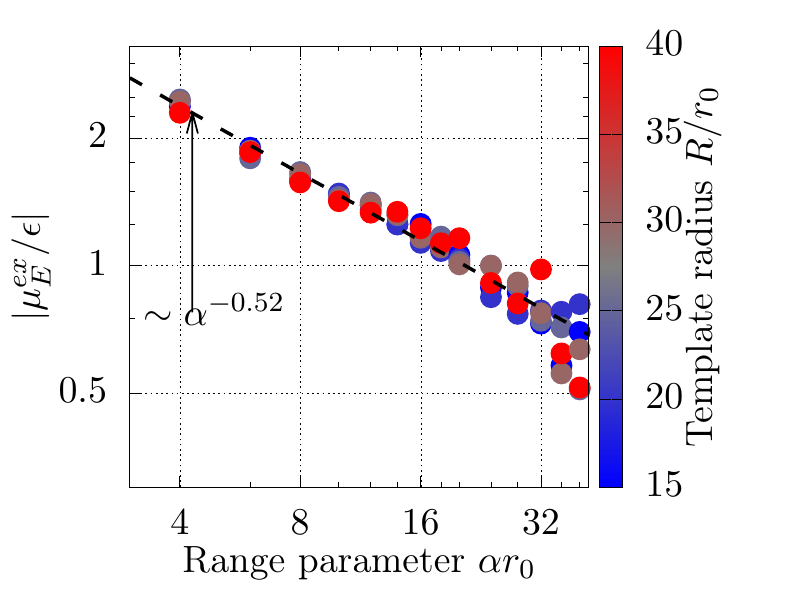}
      \caption{\label{fig:scaling-alpha-loglog}}
    \end{subfigure}
    \caption{Calculated excess chemical potential for $N$ Morse particles in linear (a) and log-log (b) scale.
      Note that increasing $\alpha$ leads to a higher chemical potential, and the scaling appears to be independent of the spherical template radius.
      \label{fig:chemical-potentials}}
  \end{center}
\end{figure}

We apply Bennett's acceptance criterion \cite{bennett-1976} with energy offset $C = 2.4\epsilon$ and sample
\begin{equation}
  e^{-\beta(F_1 - F_0 - C)} = \frac{\left<f(\beta(U_1 - U_0 - C))\right>_0}{\left<f(\beta(U_0 - U_1 + C))\right>_1}
\end{equation}
where $\beta$ is the reciprocal thermal energy $1/k_B T,$ $f(x)$ is the Fermi function $f(x) = 1 / (1 + \exp(x)),$ and the subscripts $0$ and $1$ mean that the average is obtained by sampling the potential energy function $U_0$ and $U_1,$ respectively.
The value of $C$ corresponds roughly to the average potential energy difference between the two potential energy functions, $C = \left< U_1 \right>_1 - \left<U_0\right>_0,$ which leads to a reasonably good statistics and leads to i.
We determined this value for $\alpha = 16/r_0$ but applied it for all $\alpha.$
For varying $\alpha$ the averages $\left< U_1 \right>$ and $\left< U_0 \right>$ do differ significantly, but the method remains usable for this one fixed value of $C.$

Applying Bennett's acceptance ratio means that we generate trajectories corresponding to both $U_0$ and $U_1,$ and average $f(\beta(U_1 - U_0 - C))$ over the trajectory generated by $U_0,$ while we average $f(\beta(U_0 - U_1 + C))$ over the trajectory generated by $U_1.$
This can then be converted straightforwardly into an estimate for the free energy difference between the Morse and the Morse-Einstein crystal.
The excess chemical potential of the Einstein particle $\mu_{E}^{ex}$ can be calculated analytically from the partition function and is $\mu_E^{ex} = k_BT \ln(\kappa A/2 \pi N k_BT),$ where $\kappa$ is the spring constant, $A$ is the spherical template area, $N$ the number of particles on the template and $k_BT$ the thermal energy.
Hence, the free energy difference can directly be converted into an excess chemical potential by adding $\mu_{E}^{ex},$ which we present in Fig. \ref{fig:chemical-potentials}.
The free energy difference appears to scale as $\mu_{E}^{ex} \sim \alpha^{-0.52},$ independent of the spherical template radius.

\section{Extracting the largest domain size}
\label{sec:si3}
\begin{figure}[htb!]
  \begin{center}
    \includegraphics[width=0.8\textwidth]{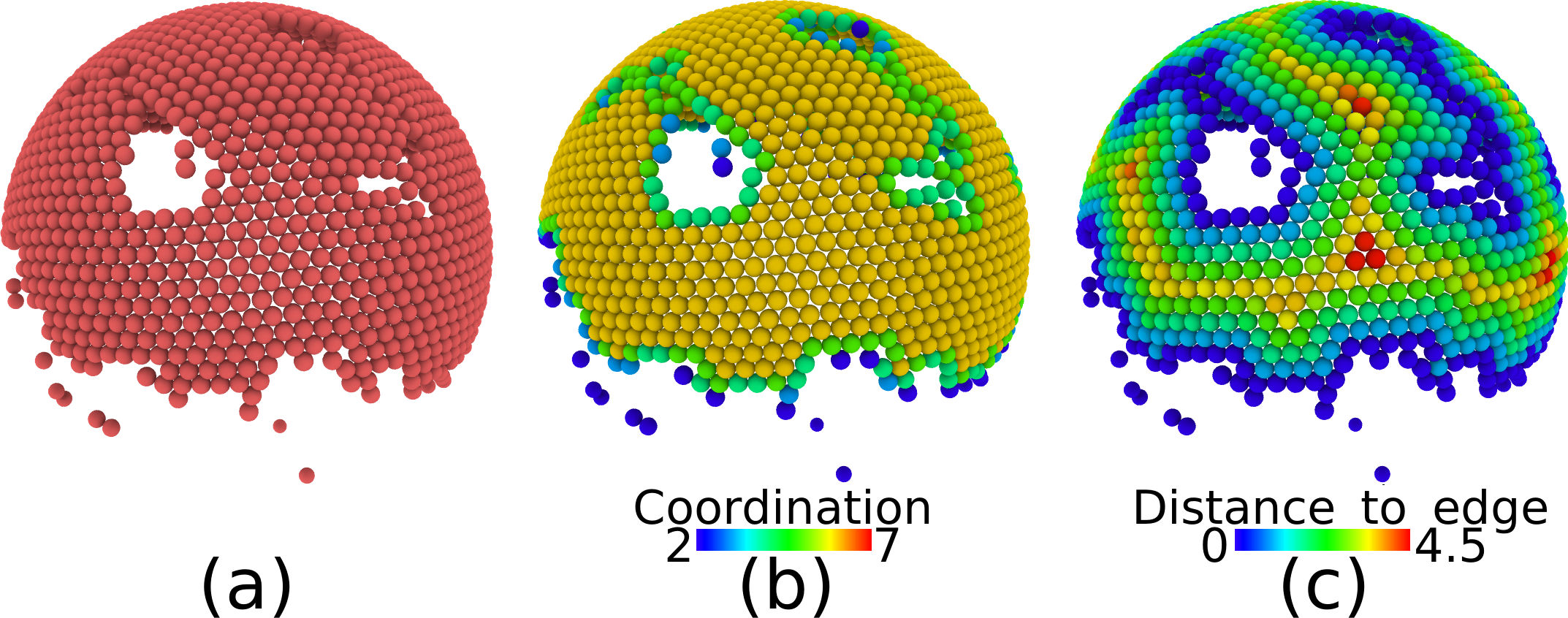}
    \caption{Different analysis stages of the same snapshot (a) obtained for $\alpha = 20/r_0, R = 15r_0$ at area coverage $\phi = 0.4.$
      (b) The coordination $n_c$ (number of nearest neighbours) identifies bulk particles ($n_c = 6$) and edge particles ($n_c \neq 6$).
      (c) We calculate for each bulk particle the shortest distance to the edge particles (colour coded, units of $r_0$).
      \label{fig:analysis-visual}}
  \end{center}
\end{figure}
To analyse the ribbon widths a series of steps is required. Each of them is explained in detail here.
First, we construct a network of nearest neighbours.
Particles are considered nearest neighbours if their distance is less than $1.3r_0$ apart. This distance coincides with the minimum after the first peak in the pair correlation function.
From this information we can identify ``edge particles'', i.e., particles that do not have six nearest neighbours.

With this information, we can compute for all particles that are not edge particles the distance to the nearest edge particles.
The largest of these distances we take as an estimate for the largest circular domain size, as it represents the largest possible circle diameter that fits inside the crystal.
This assumption breaks down for $\alpha$ sufficiently small to allow for incorporation of defects inside the lattice, rather than at the edge. This is observed for $\alpha \leq 6$ in Fig. 3(a).
In Fig. \ref{fig:analysis-visual} we present snapshots of the different stages of analysis.
With the aforementioned steps, we obtain a value for the largest circular domain diameter for each combination of spherical template radius $R$ and potential range parameter $\alpha.$

\section{Classical nucleation theory on curved surfaces}
\label{sec:theory4}

\subsection{Flat plane}
Macroscopically, the formation of a two-dimensional crystal on a \emph{flat} surface is described by Classical Nucleation Theory (CNT). \cite{kashchiev-boek}
Although this theory is well-known, we introduce it here because we later extend it with elastic terms to describe crystal formation on curved surfaces.
Furthermore, it will serve as the basis for our \emph{two-stage} model for ribbon growth, which we introduce in Section \ref{sec:theory42}.
In CNT the thermodynamic driving force towards crystal formation is opposed by a surface tension that in two dimensions is a line tension.
If we assume the line tension to be invariant to the locally exposed crystal plane, the free energy of a crystal nucleus is given by $\Delta G = N \Delta \mu  + \gamma L,$
where $N$ is the number of particles in the crystal, $\Delta \mu<0$ the chemical potential difference between the crystal phase and the surrounding liquid or gas, $L$ the crystal circumference and $\gamma$ the line tension \cite{kashchiev-boek,wu-1996}.

If we assume a circular crystal, we have $L = \pi a,$ with $a$ the diameter of the crystal so that $\Delta G_c = N \Delta \mu  + \pi a \gamma.$
Note that actual two-dimensional crystals tend to form hexagonal nuclei instead because the underlying interaction potential encourages hexagonal bond order that induces anisotropic growth. \cite{chernov-1974,kashchiev-boek,wu-1996}
However, both a hexagon and a circle exhibit the same scaling of area and circumference with the number of particles and for the purpose of extracting scaling laws it is a good approximation.
Limitless growth of the drop occurs if $\Delta G < 0,$ \emph{i.e.}, if $N > -\pi a \gamma / \Delta \mu.$
Of course, $a = 2\sqrt{N/\pi \rho}$ depends on $N$ through the equilibrium particle surface density $\rho$ of the crystal.
In other words, limitless growth occurs for $N> 4 \pi \gamma^2/(\Delta \mu)^2 \rho.$
Converting the expression for the free energy to the circle diameter $a$ leads to
\begin{equation}
  \Delta G_c = \rho \Delta \mu \pi a^2/4 + \gamma \pi a.
  \label{eqn:dG-circle-flat}
\end{equation}

Ribbon-like crystals are better described by a rectangle with a circumference $A = 2(w+l)$ and a particle number $N = \rho wl,$ leading to
\begin{equation}
  \Delta G_r = \rho \Delta \mu w l + 2\gamma (w+l).
  \label{eqn:dG-ribbon-flat}
\end{equation}
Unlike the circle, the ribbon has two parameters $w$ and $l,$ and the optimal free energy has $w = l,$ as this gives the largest area for a given circumference.
In the flat plane, a circle-like shape will always be preferred over a ribbon-like shape.
This can be seen by comparing Eqs. \eqref{eqn:dG-circle-flat} and \eqref{eqn:dG-ribbon-flat} under the constraint that they have equal areas, $wl = \pi a^2/4.$
Their difference is then simply the difference of the line tension terms, $\gamma( \pi a - 2 (w + \pi a^2/4w) ),$ which, if we substitute the optimal $w = \sqrt{\pi}a/2,$ is always negative.
Therefore, not surprisingly, in a flat plane, a ribbon is always destabilised with respect to a circle.
However, as becomes clear later, this trivial conclusion no longer holds on a curved surface.

To see whether or not a ribbon is more likely to be \emph{nucleated} than a circle, we consider the kinetic barrier heights.
For a circle, the maximum in $\Delta G_c$ corresponds to a critical diameter $a = -2\gamma/\rho\Delta\mu$ and is equal to $-\pi \gamma^2/\rho \Delta \mu,$ where we recall that $\Delta \mu < 0.$
For the ribbon the maximum is located at $l = w = -2\gamma/\rho \Delta \mu$ and has a value of $-4\gamma^2/\rho \Delta \mu.$
Therefore, in the flat plane, a ribbon-like crystal is also less likely to nucleate, as evidenced by a higher nucleation barrier in the free energy.
This means that in a two-dimensional plane a circular crystal will always be favoured over a ribbon-shaped one, both thermodynamically and kinetically.

\subsection{On a sphere}
The above picture changes if we consider crystals constrained to a spherical surface of radius $R.$
In this case, an additional energy enters the free energy that takes into account the elastic cost of bending the crystal to accommodate the curved template. \cite{schneider-2005,majidi-2008}
First consider again the circular crystal.
To make the analysis more straightforward, we scale the free energy to the spherical surface area, $4 \pi R^2,$ times Young's modulus $Y.$
This leads to a reduced unit $\eta := a/R$ and a dimensionless free energy \cite{schneider-2005,majidi-2008,meng-2014,koehler-2016}
\begin{equation}
  \Delta g_c = \frac{1}{4 \pi }\left[ \frac{\rho \Delta\mu \pi}{4 Y}\eta^2 + \frac{\pi \gamma}{RY} \eta + \frac{\pi}{24576}\eta^6 \right].
  \label{eqn:dG-circle-sphere}
\end{equation}
Although the elastic term does not influence the kinetic barrier height or critical diameter significantly \cite{gomez-2015}, it does influence the thermodynamic stability of the circular crystal.
For very large $Y,$ the elastic term dominates to such an extent that the local minimum in the free energy that occurs for sizes larger than the critical nucleus size becomes larger than $0,$ as is pointed out in Refs. \cite{vitelli-2006,gomez-2015,koehler-2016}.
In that case, $\Delta g_c \geq 0,$ and the formation of a circular crystal is thermodynamically suppressed.

Furthermore, for $Y > 0,$ there is now an optimal circular diameter that minimises the free energy.
Under the assumption $\gamma \ll R Y$ this minimum is located at $\eta = (-6144 \rho \Delta \mu / Y)^{1/4} := \eta_0.$
In Refs. \cite{meng-2014,koehler-2016} this value is associated with the critical circular crystal size at which growth transitions to a ribbon, as circular domains larger than this radius have a higher free energy.
This scaling we shall refer to as the \emph{optimal circle scaling}.
However, as long as $\Delta g_c<0,$ the circle can in principle continue to grow, for example if there is not a sufficient amount of material to nucleate a second, stable circular cluster.

Unlike in the flat plane, on a sphere a ribbon-shaped crystal can be more stable than a circle due to a different scaling of the elastic energy with the crystal dimensions.
The dimensionless free energy for the ribbon on a spherical surface is given in terms of dimensionless width $\omega = w/R$ and length $\lambda = l/R$ by \cite{schneider-2005,majidi-2008}
\begin{equation}
  \Delta g_r = \frac{1}{4 \pi}\left[ \frac{\rho \Delta \mu}{Y} \omega \lambda + \frac{2 \gamma}{RY}( \omega + \lambda ) + \frac{\omega^5\lambda}{640(1-\nu^2)} \right],
\end{equation}
where $\nu$ is the Poisson ratio of the material.
Note that in this expression, $\rho \Delta \mu \lambda + 2\gamma/R$ has to be negative, and hence it only holds for $\rho \Delta \mu < -2\gamma/R\lambda = -2\gamma/L.$


The reason a ribbon can be stabilised over a circular crystal on a curved surface is because although the elastic strain scales with $\omega^5,$ it only scales linearly with $\lambda.$
This leads to a different scaling with the total crystal area, which becomes obvious when the free energy of the ribbon is expressed in terms of the width $\omega$ and dimensionless area $A/R^2 := \zeta = \omega \lambda,$
\begin{equation}
  \Delta g_r = \frac{1}{4 \pi}\left[ \left(\frac{\rho \Delta \mu}{Y}  + \frac{\gamma}{RY} \frac{2}{\omega} + \frac{\omega^4}{640(1-\nu^2)}\right)\zeta  + \frac{\gamma}{RY}2 \omega \right].
  \label{eqn:dgr-area}
\end{equation}
This free energy can in principle be optimised with respect to $\omega$ to obtain an optimal ribbon width for a given area:
\begin{equation*}
  4\pi\frac{\partial \Delta g_r}{\partial \omega} = \left[ \frac{\rho\Delta\mu}{Y} -\frac{\gamma}{RY}\frac{2}{\omega^2} + \frac{\omega^3}{160 (1-\nu^2)}\right] \zeta + 2\frac{\gamma}{RY} = 0
\end{equation*}
Because the lowest order term $-2\gamma/RY \zeta \omega^2$ is negative, there will always be some $\omega$ that optimises Eq. \eqref{eqn:dgr-area}.

For the ribbon to be stable at some point, the total free energy should be negative, i.e., $\Delta g_r < 0.$
Since $2 \omega \gamma/RY>0,$ the ribbon can only be stable when
\begin{equation*}
  \left[ \left(\frac{\rho \Delta \mu}{Y}  + \frac{\gamma}{RY} \frac{2}{\omega} + \frac{\omega^4}{640(1-\nu^2)}\right)\right]\zeta < -2\omega \frac{\gamma}{RY}.
\end{equation*}
Since $0 \leq \zeta \leq 1,$ this condition can only be satisfied if at least
\begin{equation*}
\frac{\rho \Delta \mu}{Y}  + \frac{\gamma}{RY} \frac{2}{\omega} + \frac{\omega^4}{640(1-\nu^2)} + 2\omega \frac{\gamma}{RY} \leq 0.
\end{equation*}
The equality can only be achieved for sufficiently small $\rho \Delta \mu/Y$ and $gRY,$ i.e., at sufficiently large Young's moduli.
When this inequality holds, a ribbon can in principle be indefinitely large and still have a negative free energy, unlike a circular crystal.

However, the kinetic barrier associated with a ribbon is always larger than that of a circle.
Under the assumption that the elastic contribution is negligible for crystals smaller than the critical nucleus size, we find that the barrier height, i.e., the maximum in $\Delta g$, is located at $\eta_b = -2\gamma/\rho\Delta\mu R$ for the circle and at $\omega_b = \eta_b, \zeta_b = \omega^2_b$ for the ribbon.
The corresponding free energies are $\Delta g_c^b = -\gamma^2/4 \rho\Delta\mu R^2$ and $\Delta g_r^b =  -\gamma^2/\pi \rho \Delta\mu R^2.$
Hence, the kinetic barrier for formation of a circle is lower than that of a ribbon by a factor of $\pi/4.$

This implies that, while under certain conditions a ribbon-shaped crystal has a lower free energy, it is less likely to be nucleated.
A more likely scenario that leads to the formation of ribbon-like structures is therefore a \emph{two-stage} nucleation, in which the ribbon grows out of a circular nucleus.
In such an event, the crystal has the kinetic energy barrier of a circlular shape and initially grows as such.
However, at some point, the free energy decrease for continued growth as a circle is smaller than that for growth as a ribbon, and hence, the crystal continues growth into ribbon-like structures.
Such a growth pathway is also more consistent with the experimental results \cite{meng-2014} and crystal phase field calculations. \cite{koehler-2016}
In Section \ref{sec:theory42} we present a free energy for such a growth type.

\subsection{Two-stage nucleation theory on a sphere}
\label{sec:theory42}

In Section \ref{sec:theory4} we argue that the free energy of a circular crystal is, under certain conditions, lower than that of a ribbon.
Hence, the formation of a pure ribbon-like crystal is not probable, as ribbons smaller than the transition size are destabilised with respect to circular crystals with the same area.
This motivates us to investigate the possibility of a two-stage nucleation, meaning that a ribbon-shaped structure grows out of a pre-existing, circular nucleus.
For this to happen, the free energy gain due to a small area increment in the shape of a ribbon should be smaller than that due to an equal area increment in the shape of the already existing circle.
Both growth modes are sketched in Fig. \ref{fig:increments}.

\begin{figure}[tb]
  \begin{center}
    \includegraphics[width=0.35\textwidth]{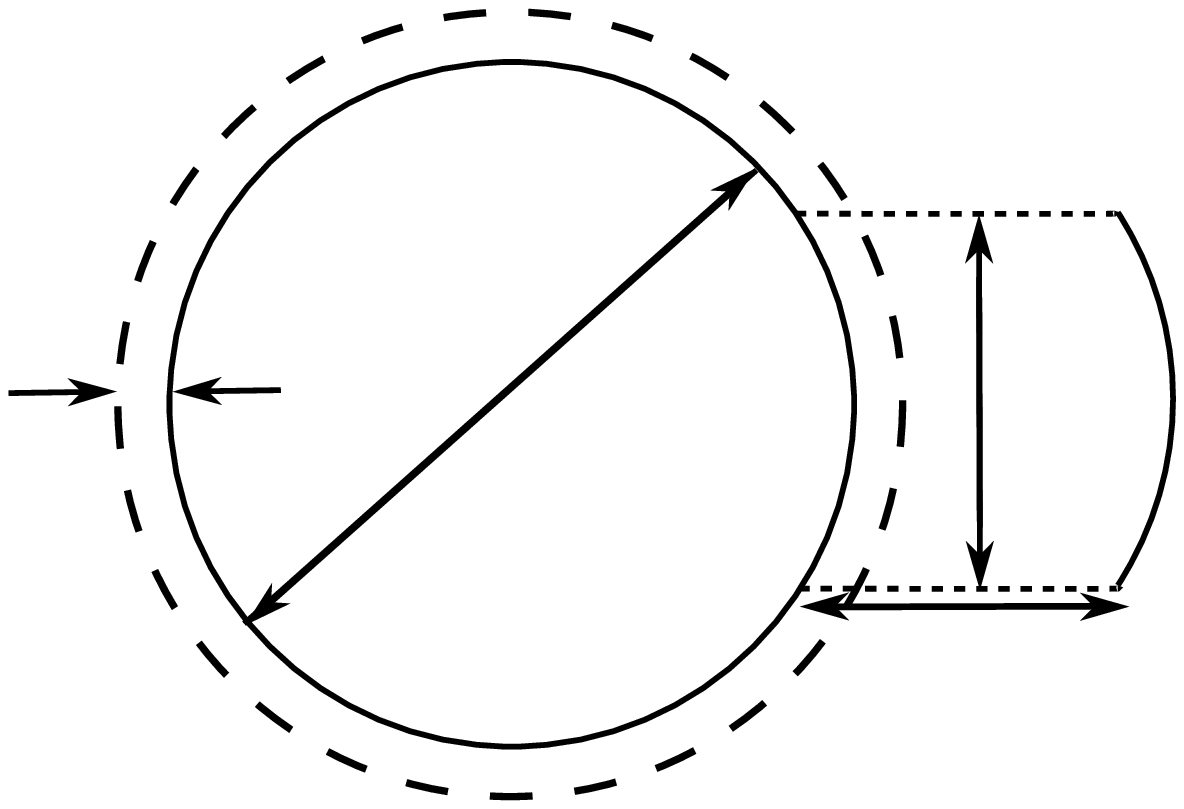}
    \caption{Two possible modes of growth from a circular nucleus of diameter $a.$ Either growth continues along an infinitesimal diameter increase $\dd{a},$ or as a ribbon-like structure of width $w$ and length $\dd{l}$ protruding out of the circular nucleus. \label{fig:increments}}
  \end{center}
\end{figure}

From the typical crystal morphologies presented in Refs. \cite{meng-2014} and \cite{koehler-2016}, it appears that the ribbons grow out of the initial crystals without distorting the lattice vectors.
Therefore, we assume there is no line tension associated with the circle-ribbon interface.
We furthermore assume that the line tension associated with the width of the ribbon is of the same as the previously exposed circular rim that is now covered by the ribbon.
This means there is no line tension term associated with the width, because it is already accounted for by the original line tension term.
With these assumptions, the total free energy of the hybrid crystal becomes 
\begin{equation}
  \Delta g_{c+r} = \frac{1}{4\pi}\left[ \frac{\rho \Delta \mu}{Y}\left( A_c + A_r  \right) + \frac{\gamma}{RY} \left( \sqrt{4\pi A_c} + 2 \frac{A_r}{\omega} \right) + \frac{A_c^3}{384 \pi^2} + \frac{\omega^4 A_r}{640(1-\nu^2)} \right],
  \label{eqn:hybrid-crystal-dG}
\end{equation}
where $A_r$ is the dimensionless ribbon area $\omega \lambda R^2$ and $A_c$ the circle area $\pi (R \eta)^2/4.$
The free energy change of the crystal $\dd{\Delta g_{r+c}}$ of growing purely as a circle can be determined by expanding $\Delta g_{c+r}$ in a Taylor series around $A_c$ for $A_r = 0,$ leading to
\begin{equation}
  \dd{\Delta g_{c+r}} = \left(\frac{\partial \Delta g_{c+r} }{\partial A_c} \right)\dd{A} = \frac{\dd{A}}{4\pi R^2}\left[ \frac{\rho \Delta \mu}{Y} + \frac{\gamma}{RY}\sqrt{\frac{\pi}{A_c}} + \frac{A_c^2}{128 \pi^2} \right].
  \label{eqn:hybrid-crystal-dG-incr-eta}
\end{equation}
To determine the free energy change of growing as a ribbon, we expand $\Delta g_{c+r}$ in terms of $A_r$ for fixed $\omega$ and $A_c,$ leading to 
\begin{equation}
  \dd{\Delta g_{c+r}} = \left(\frac{\partial \Delta g_{c+r} }{\partial A_r}\right) \dd{A} = \frac{\dd{A}}{4 \pi R^2}\left[ \frac{\rho \Delta \mu}{Y} + \frac{2 \gamma}{RY \omega} + \frac{\omega^4}{640(1-\nu^2)} \right].
  \label{eqn:hybrid-crystal-dG-incr-lambda}
\end{equation}
The width that gives the optimal free energy change follows from optimising $\dd{\Delta g_{c+r}}$ with respect to $\omega,$ which leads to $\omega = [320(1-\nu^2)\gamma / RY]^{1/5}.$

In other words, the free energy change due to continued growth as a circle from an area $A_c$ with increment $\Delta A$ is given by $( \partial \Delta g_{c+r} / \partial A_c) \Delta A$ and the free energy change due to growth as a ribbon out of a circular nucleus is given by $( \partial \Delta g_{c+r} / \partial A_r) \Delta A.$
Therefore, the difference between Eqs. \eqref{eqn:hybrid-crystal-dG-incr-eta} and \eqref{eqn:hybrid-crystal-dG-incr-lambda}, $\Delta \Delta g,$ indicates whether continued growth favours a ribbon or a circle,
\begin{align}
  \Delta \Delta g :=& \left( \frac{\partial \Delta g_{c+r}}{\partial A_c} - \frac{\partial \Delta g_{c+r}}{\partial A_r} \right) \Delta A \notag \\
  =& \frac{\Delta A}{4\pi R^2}\left[ \frac{\gamma}{RY}\left( \sqrt{\frac{\pi}{A_c}} - \frac{2}{\omega} \right) + \frac{A_c^2}{128 \pi^2} - \frac{\omega^4}{640(1-\nu^2)} \right].
  \label{eqn:dg-diff}
\end{align}

Continued growth as a ribbon is preferred under two conditions. First of all, it is necessary that $\Delta \Delta g > 0,$ as only then the free energy gain for growing as a ribbon is preferred over growing as a circle.
Secondly, it is required that $\Delta g_{c+r} < 0,$ as else the formation of any type of crystal at all is destabilised.
Finally, although not technically a necessity for the initial onset of ribbon formation, we check whether or not $\partial \Delta  g/\partial A_r<0.$
If this is the case, the crystal can grow indefinitely as a ribbon, since continued growth as a ribbon will lead to a further decrease in the free energy.

To determine the critical area at which the transition takes place, we first numerically determine for a range of $\gamma / RY$ and $\rho \Delta \mu$ the roots of $\Delta \Delta g,$ i.e., the points that satisfy the first criterion.
Then, we determine whether or not $\Delta g_{c+r} < 0$ and what the sign of $\partial \Delta g_{c+r}/\partial A_r$ is.
We find that $\Delta g_{c+r} < 0$ is only satistfied at the roots of Eq. \eqref{eqn:dg-diff} for significantly large ratios for $|\rho \Delta \mu R / \gamma|.$
In particular, we find that at a ratio of at least $6$ is required to obtain a transition at an area $A_c < 0.6$ that satisfies all criteria, corresponding to $\gamma/RY = 2\times 10^{-4}$ and $\rho \Delta \mu/Y = 12\times 10^{-4}.$
Furthermore, although increasing Young's modulus leads to a decrease in the critial area at which $\Delta \Delta g = 0,$ it also leads to a higher free energy, occasionally such that $\Delta g_{c+r} > 0$ at the transition.
Hence, although larger Young's moduli favour ribbon formation, they will only actually form at a proportional decrease in the chemical potential.

To determine how the elastic cost influences the area at which the transition occurs, we determine how the critical circular area scales with Young's modulus for various ratios of the chemical potential and line tension terms.
In figure \ref{fig:critical-scaling} we plot our findings, which clearly show that the critical area scales as $A_c \sim (\gamma/RY)^{2/5}$ and hence the diameter of the largest circular domain size scales as $\eta \sim (\gamma/RY)^{1/5}.$
Note that all points collapse on the same curve given by $A_c = 23.6594(\gamma/RY)^{0.4}.$
The relative standard errors in the linear regression are $1.3 \times 10^{-4}\%$ for the prefactor and $3.5\times 10^{-5}\%$ for the exponent.
Both the prefactor and scaling exponent are independent of $\rho \Delta \mu / Y.$

We notice a clear breakdown in this scaling for $\gamma/RY > 2 \times 10^{-4}.$
In this regime, the line tension is prohibitively large and suppresses ribbon formation.
There is another breakdown for sufficiently small $\gamma/RY$ in combination with small $\rho \Delta \mu/Y.$
In this regime, the chemical potential is not negative enough to compensate for the combined costs of bending and the line tension, in other words, there is no negative free energy $\Delta g_{c+r} < 0$ anymore.

\begin{figure}[tb]
  \begin{center}
    \begin{subfigure}{0.48\textwidth}
      \includegraphics[width=\textwidth]{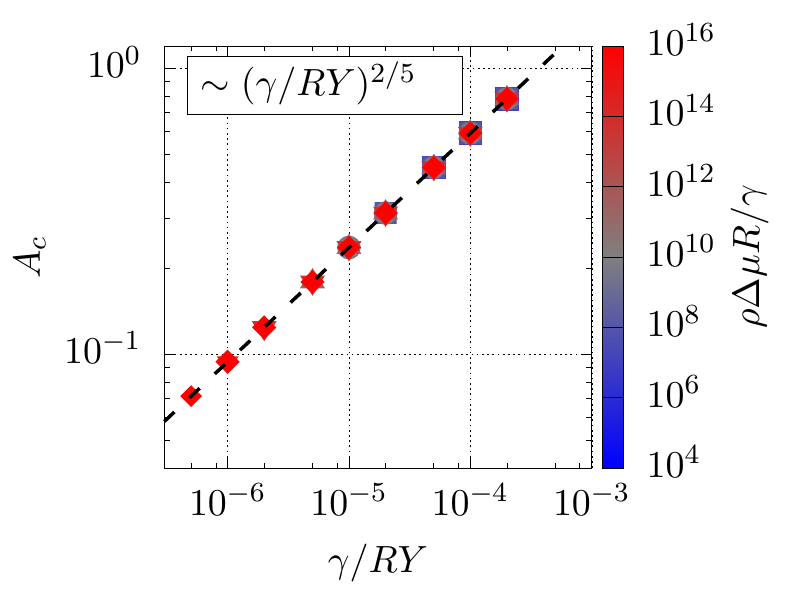}
    \end{subfigure}
    \begin{subfigure}{0.48\textwidth}
      \includegraphics[width=\textwidth]{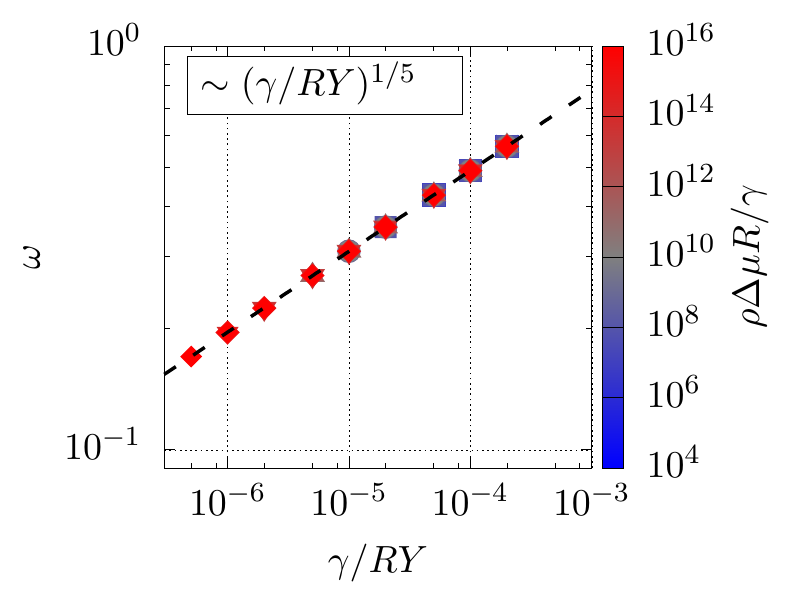}
    \end{subfigure}
    \caption{Scaling of the area of the circular domain (left) and smallest ribbon width (right) at which growth transitions from a circular crystal to a ribbon as a function of the ratio of the dimensionless chemical potential $\rho \Delta \mu$ and line tension $\gamma/RY.$
      Note that all points collapse onto the same curve.
      For smaller $\rho \Delta \mu R / \gamma$ there is no ribbon formation for smaller $\gamma / RY$ because the free energy is no longer negative.
      \label{fig:critical-scaling}}
  \end{center}
\end{figure}

Hence, our analysis suggests that, if the formation of ribbons indeed proceeds as a two-stage, heterogeneous nucleation of a ribbon-like structure on a pre-existing, circular nucleus, the normalised area of this circular nucleus should scale as $A_c = \pi a^2/16 \pi R^2 \sim (\gamma/RY)^{2/5},$ where $a$ is the diameter of the circular crystal and $R$ the radius of the template.
This is different from the scalings previously proposed in Refs. \cite{meng-2014,koehler-2016}, in which the transition area is associated with the minimum in $\Delta g_c,$ which, in the limit of negligible line tension, scales as $A_c \sim (\rho \Delta \mu / Y)^{1/4}.$

The actual scaling of the transition area can be obtained from computer simulations by determining the largest circular domain size as a function of the spherical template radius, the chemical potential, the line tension and Young's modulus, as is described in the main text and in the previous sections.

\bibliographystyle{/home/stefan/mystyle}

\end{document}